\documentclass[journal=langd5,manuscript=article]{achemso}
\DeclareSymbolFont{matha}{OML}{txmi}{m}{it}% txfonts
\DeclareMathSymbol{\varv}{\mathord}{matha}{118}
\usepackage[version=3]{mhchem} % Formula subscripts using \ce{}
\usepackage{amssymb}
\usepackage{bm}
\usepackage{xcolor}
\usepackage{soul,ulem}
\usepackage{physics}
\usepackage{upgreek,textgreek}
\usepackage{graphicx}%
\usepackage{multirow}%
\usepackage{amsmath,amssymb,amsfonts}%
\usepackage{amsthm}%
\usepackage{mathrsfs}%
\usepackage[title]{appendix}%
\usepackage{xcolor}%
\usepackage{textcomp}%
\usepackage{manyfoot}%
\usepackage{booktabs}%
\usepackage{algorithm}%
\usepackage{algorithmicx}%
\usepackage{algpseudocode}%
\usepackage{listings}%
\author{Lu\'is H. Carnevale}
\affiliation[ifpan]
{Institute of Physics, Polish Academy of Sciences, Al. Lotnik\'ow 32/46, 02-668 Warsaw, Poland}
\email{carnevale@ifpan.edu.pl}
\author{Gabriela Niechwiadowicz}
\affiliation[ifpan]
{Institute of Physics, Polish Academy of Sciences, Al. Lotnik\'ow 32/46, 02-668 Warsaw, Poland}
\author{Panagiotis E. Theodorakis}
\affiliation{Institute of Physics, Polish Academy of Sciences, Al. Lotnik\'ow 32/46, 02-668 Warsaw, Poland}

   \title[Coarse-Grained Model of Sodium Dodecyl Sulfate Anionic Surfactant based on the MDPD-Martini Force-Field]
  {Coarse-Grained Model of Sodium Dodecyl Sulfate Anionic Surfactant based on the MDPD-Martini Force-Field}


\begin{document}

\begin{tocentry}
\includegraphics[height=4.45cm]{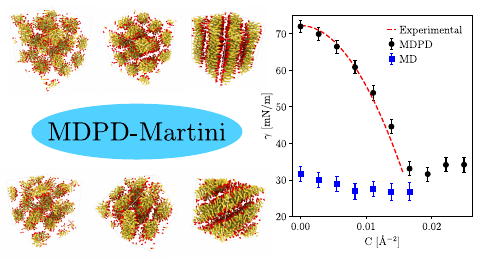}
\end{tocentry}

\begin{abstract}
Sodium dodecyl sulfate (SDS) surfactant is widely used in various 
applications, such as household products (e.g., shampoos, toothpaste, 
detergents, cleaning products) and food manufacturing (e.g., emulsifier).
To investigate its properties via computer simulation, 
various models have been developed,
including coarse-grained (CG) ones that are suitable for capturing
surfactant's self-assembly, as well as fundamental properties for aqueous 
systems with surfactant, such as surface tension. 
Here, we present a CG model for SDS--water systems
for many-body dissipative particle dynamics (MDPD), which is based
on the MDPD-Martini force-field (FF). 
In the model, charged groups, namely the SDS sulfate head group and the
sodium cation, are explicitly modeled following
the standard mapping of the Martini force-field for molecular dynamics (MD),
while the rest of interactions have been obtained from previous
MDPD-Martini models for lipid systems, thus demonstrating their transferability. 
Various relevant system properties, such as the coherent scattered
intensity and surfactant distribution at the liquid--vapor
surface, are investigated and results 
are compared to those obtained by MD simulations and experiments
for different surfactant concentrations. 
Our findings indicate that MDPD-Martini models can offer a credible 
alternative to MD-Martini models, also, for systems with explicit charges as 
shown here for SDS. Moreover, MDPD-Martini models reproduce nicely 
the experimental surface-tension isotherm, in contrast to MD simulations.
In view of the transferability of the MDPD-Martini interactions,
the model parameters of this study can be tested and used to simulate 
a wider range of soft-matter systems.
\end{abstract}

\vspace{0.7in}

\section{INTRODUCTION}

Surfactants are commonly used in various applications
as detergents \cite{Broze1999}, emulsifiers \cite{Becher1965,Goodarzi2019},
wetting and foaming agents \cite{Rosen2012,Blute1994}, etc. 
The broad use of surfactants in applications is due to
their amphiphilic nature, that is
they consist of a hydrophilic and a hydrophobic part. For this reason,
they can favorably adsorb at interfaces, where they can
reduce the surface tension, \cite{Lunkenheimer2003} thus alleviating the
tension between different phases. Moreover, at
concentrations above a critical aggregation concentration (CAC), 
they can form aggregates of different size and shape, for example
spherical micelles, etc. \cite{Rosen2012} and therefore various properties of 
aqueous systems with surfactant highly depend on surfactant concentration.
Moreover, surfactants can be nonionic, such as
alkyl ethers \cite{Berthod2001,Arbabi_2023_soft_matter,Theodorakis2015Langmuir},
ionic, including both anionic \cite{Missel1980,Cooper1963} 
and cationic \cite{Boethling1992} surfactants,
and zwitterionic surfactants that contain both
positive and negative charged groups \cite{FernLey1978}. 

Sodium dodecyl sulfate (SDS) is an anionic surfactant that is
widely used in industry. For example, it is commonly used in
commercial products such as shampoos and cleansers.
For this reason, understanding its properties has been
the focus of both experimental and theoretical research alike,
in particular properties such as the size and shape of SDS
micelles and determining the CAC above which these occur \cite{Cabane1985,Bergstrom1999,Magid2000,Kakitani1995,Ludwig2021,Khodaparast2021,Bezzobotnov1988,Gawali2019,Lipfert2007,Aniansson1976,Quina1995,Dutkiewicz2002,Khan2019}.
At the same time, molecular simulation has been an important tool for
investigating aqueous systems with SDS surfactant, since it
can often provide insights beyond experimental capabilities
by tracking each molecule at all times during the simulation.
Still, simulating systems with
surfactant is challenging even for molecular dynamics (MD) simulations
of coarse-grained (CG) models, since the time scale involved to fully
capture certain phenomena, such as the self-assembly and diffusion of
surfactant aggregates, would require adequately long simulations or following simulation
protocols (e.g. simulation annealing methods\cite{Yatsyshin2020}) for efficient
sampling, for example, to overcome metastable energy minima toward a global minimum.
Despite previous MD studies, 
which have provided valuable insights into the properties of water--SDS 
systems \cite{Peroukidis2021,Tang2014,Jalili2009,Palazzesi2011,Chun2015,Anogiannakis2020,Zhou2021,Wang2012}, 
many challenges still persist, such as reliable surface-tension measurements. For this reason,
there has been a quest for improving CG models, including
the further reduction of associated computational costs of
these methods.
In this regard, dissipative particle dynamics (DPD) \cite{Anderson2018,Mao2015,Gray2023} 
and many-body dissipative particle dynamics 
(MDPD) \cite{Hendrikse2024,Ghoufi2013,Zhou2019,Carnevale2024_phys_fluids,Hendrikse2023} simulation
methods that rely on soft, short-range interactions have 
emerged as a suitable alternative to MD simulation models that usually
rely on hard-core interactions (e.g. Lennard-Jones).  Moreover, the
sodium ion has thus far mostly been part of the hydrophilic head of the SDS 
surfactant, thus represented by an uncharged interaction 
site in MDPD models \cite{Ghoufi2013,Zhou2019,Carnevale2024_phys_fluids}.
However, it has recently been shown that sodium ions can actually
be represented as separate point charges \cite{Hendrikse2024} 
as in various all-atom and CG models
commonly employed in MD simulations (e.g. Martini \cite{Marrink2004,Souza2021}),
instead of using a smeared charge approach due to the soft nature of the 
potential in DPD and MDPD models \cite{Anderson2018,Mao2015,Gray2023}. 
Furthermore, avoiding the so-called charge collapse related to the presence of 
long-range interactions due to the presence of point charges in the simulations
has opened new possibilities for developing MDPD force-fields that rely on 
point charges as separate beads,\cite{Hendrikse2024} a feature that is 
indispensable for developing a general-purpose force-field (FF).
Hence, this specific aspect is key for the MDPD-Martini 
models,\cite{Carnevale2024_MDPD_MARTINI,Kramarz2025}
which will be here presented for aqueous systems with SDS surfactant.

The MDPD-Martini force-field (FF) \cite{Carnevale2024_MDPD_MARTINI,Kramarz2025}
is generally based on the Martini ``LEGO'' approach and its
mapping \cite{Marrink2004,Marrink2007,Marrink2013,Alessandri2021} for describing
the interactions between the different bead types. 
It has been used for different lipid 
systems offering a significant speed-up in simulations 
with respect to those relying on
MD-Martini models without compromising in the quality of 
results regarding various relevant properties. Furthermore, the self-assembly process of lipid bilayers was shown to be 4-7 times less computationally expensive, depending on the system size, when comparing MDPD-Martini with MD-Martini.\cite{Carnevale2024_MDPD_MARTINI,Kramarz2025}
This is a significant advantage with respect to MD-Martini simulations,
thus offering in general the capability of simulating larger systems for longer times.
Moreover, MDPD models are able to reproduce hydrodynamic
interactions, which is generally challenging for MD simulations
and they can reach equilibrium states of soft-matter complex 
systems with greater ease than in the case of 
MD simulations.\cite{zhao2021-review}
Here, we proceed one step further and  show how the MPDD-Martini models apply 
in the case of water--SDS systems with the sodium particles being 
represented by individual point charges
following the mapping recipe of MD-Martini models.\cite{Souza2025,Marrink2007} 
A key feature of the MDPD-Martini
approach is the transferability of the interactions (``LEGO approach''),
which renders this force-field suitable for simulating a range
of different systems (general-purpose), as in the case of the standard MD
Martini. For this reason, MD-Martini models have been applied for a
wide range of systems, such as proteins \cite{Monticelli2008,Periole2009,deJong2013},
polymers \cite{Lee2009}, carbohydrates \cite{Lopez2009},
glycolipids \cite{Lopez2013}, glycans \cite{Chakraborty2021},
DNA \cite{Uusitalo2015}, RNA \cite{Uusitalo2017}, water \cite{Yesylevskyy2010}
and various solvents \cite{Vainikka2021}.
Moreover, various extensions include simulations for specific 
pH \cite{Grunewald2020}, chemical reactions
(reactive Martini) \cite{Sami2023}, and the G\={o}Martini approach
for proteins \cite{Poma2017}, which has been also integrated into 
the Martini 3.0 release.\cite{Souza2025}

In this study, we have carried out MDPD simulations
based on the MDPD-Martini-FF for water--SDS systems for various
surfactant concentrations. Several properties for these systems have been
calculated and juxtaposed with previous MD-Martini results and experiments.\cite{Anogiannakis2020,Fainerman2010,Hammouda,Peroukidis2021}
The MDPD-Martini interactions previously obtained for lipid 
systems\cite{Carnevale2024_MDPD_MARTINI,Kramarz2025} demonstrate transferability to 
the current water--SDS system. Moreover, sodium-cation, with its first hydration shell, and SDS-anionic-head 
groups are represented by distinct Qd and Qa Martini-type beads, respectively.
We find that the MDPD-Martini models perform better for properties, such as
the surface tension isotherm in comparison with the MD-Martini model
and in good agreement with the experiment.\cite{Fainerman2010} Moreover,
the self-assembly morphologies as a function of concentration agree well between 
the MD and MDPD models, as well as other properties, such as the
density distribution of surfactant at the liquid--vapor surface
and its thickness, and the coherent scattered intensity. Thus, while
our study provides the MDPD-Martini-FF parameters for SDS--water systems with
explicit charges, it also opens possibilities for simulating a wider range
of systems due to the transferability of the currently developed
MDPD-Martini-FF interaction matrix, which can be further tested in practice in future work.

\section{MATERIALS AND METHODS}

% \begin{figure}[bt!]
% \centering
% \includegraphics[width=\columnwidth]{FIGS/fig1.png}
%  \caption{
% Typical conformations 
% } \label{fig:1} 
% \end{figure}

\begin{figure}[t]
    \centering
    \includegraphics[width=.6\linewidth]{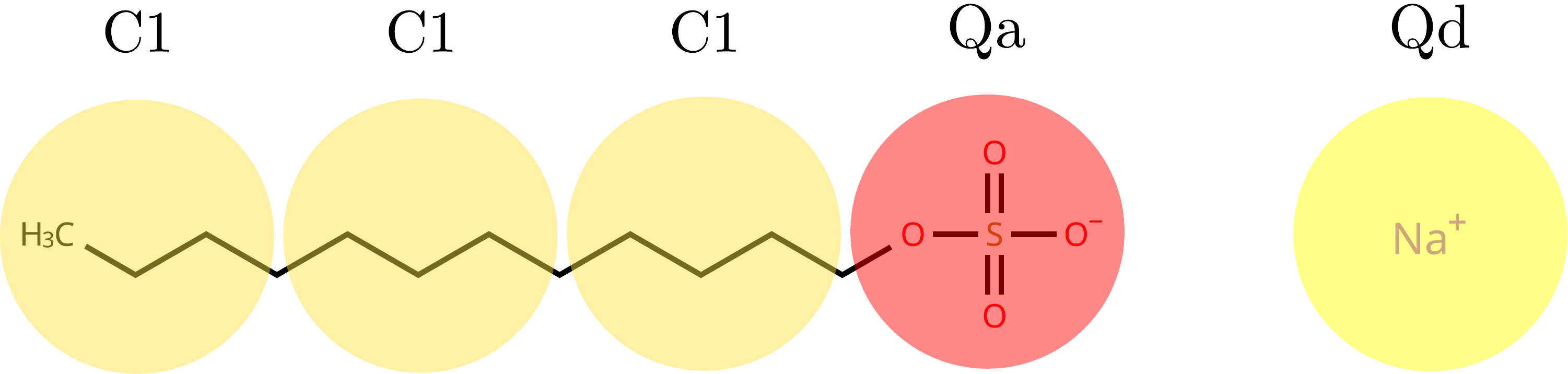}
    \caption{Coarse-grained representation of the SDS surfactant and sodium counter ion used in both MD and MDPD simulations. SDS beads are tethered together 
    with harmonic interactions (see text for details), while
    each sodium ion, with its first hydration shell, is represented by a point-charge bead, which is not
    bonded to the SDS chain.}
    \label{fig:model}
\end{figure}

\subsection{Molecular Dynamics}

Simulations for both the MD and the MDPD models are carried out
by using LAMMPS software \cite{Plimpton1995,Thompson2022}.
Following the usual Martini approach, MD simulations were run with a shifted 
Lennard--Jones potential where the energy goes to zero smoothly between an inner
and an outer cutoff, $r_{\rm in}=0.9$~nm and $r_{\rm out}=1.2$~nm, respectively.
Long-range electrostatic interactions were computed with the particle--particle
particle--mesh (pppm) method and by setting the electric permittivity 
to $80$ following previous studies.\cite{Anogiannakis2020,wang2015}
This method takes into account the electrostatic interactions explicitly through the Coulomb potential for particles up to the computational cutoff and the longer range interactions are computed from the Ewald sum on reciprocal space.
Both temperature and pressure in the simulations
were controlled through the Nos\'e--Hoover thermostat as
implemented in LAMMPS. Temperature was kept at $T=300$~K and pressure at $P=1$~atm.
A $20$ fs time-step was used in all MD simulations.

The Martini model for the SDS surfactant is illustrated in 
\textbf{Figure~\ref{fig:model}}.
It consists of the $Q_a$ head-group bead with a negative electric charge
and three $C_1$ tail beads tethered together with harmonic interactions to 
form the SDS chains. 
An equal number of sodium cations to the total number of SDS chains
is also included in the simulation box to neutralize the system.
These cations are represented by the $Q_d$ bead-type. 
All simulations were done with water as an explicit solvent modeled 
by $P_4$ Martini beads. A non-polarizable water model was used due to its much lower computational cost when compared to the Martini polarizable water model. Furthermore, most models in the literature use non-polarizable water for the SDS--water system, making it a suitable option to validate our MD simulations.

The specific Lennard-Jones interaction parameters, namely $\epsilon_{\rm ij}$ and $\sigma_{\rm ij}$,
for each pair of beads are reported in Table~\ref{tab:lj}. The harmonic potential
between bonded beads in a surfactant molecule was set to have a force constant
of $K=1250$~kJ~mol$^{-1}$~nm$^{-2}$ and bond length $r_0=0.47$~nm. A harmonic 
cosine potential between three consecutive bonded beads was also used with a 
force constant of $K=25$~kJ~mol$^{-1}$ and equilibrium angle $\theta_0=180^{\circ}$.\cite{Marrink2007,Anogiannakis2020}

\begin{table}[t]
    \centering
    \begin{tabular}{c|ccccc}
         & P4   & Qd   & Qa   &  C1   \\
    \hline
    P4   & 5.0 & 5.6 & 5.6 & 2.0  \\ 
    Qd   &     & 5.0 & 5.6 & 2.0  \\ 
    Qa   &     &     & 5.0 & 2.0  \\ 
    C1   &     &     &     & 3.5  \\ 
    \end{tabular}
    \caption{Lennard-Jones $\epsilon_{\rm ij}$ parameter for each bead interaction in kJ~mol$^{-1}$. $\rm i$ and $\rm j$ represent the different bead types. The distance, $\sigma_{\rm ij}$, was set to $0.47$~nm, except for
    Qd-C1 and Qa-C1 interactions, where it was set to $0.62$~nm.\cite{Anogiannakis2020}}
    \label{tab:lj} 
\end{table}

\subsection{Many-Body Dissipative Particle Dynamics}

The MDPD method has been applied for investigating
a range of different systems, which includes
different types of fluids \cite{Espanol1995,warren2003,zhao2017,zhao2021-review,zhao2021,Han2021,Vanya2018}
and their properties, such as surface tension \cite{Carnevale2023}.
However, it is also suitable for investigating 
multi-phase and multicomponent soft-matter systems,
such as systems with surfactants \cite{Hendrikse2024,Carnevale2024_phys_fluids}.
During a standard MDPD simulation, the Langevin equation of 
motion is solved for each particle of the system
as in MD \cite{Theodorakis2011}, but
the pair-wise forces are a direct input in the case of MDPD simulations.
The Langevin equation reads
\begin{eqnarray}
	m\frac{d\bm{v}_i}{dt} = \sum_{j\neq i} \bm{F}_{ij}^C + \bm{F}_{ij}^R + \bm{F}_{ij}^D,
	\label{eq1}
\end{eqnarray}
where $\bm{F}^C$ is the conservative force on each particle,
and $\bm{F}^R$ and $\bm{F}^D$ are the random and dissipative
forces, respectively, which act as a thermostat. 
Hence, the latter forces are related through the fluctuation--dissipation
theorem. The mass, $m$, of the beads is the same for all particles and
set to unity.

A main difference between the MDPD method and its predecessor, DPD, \cite{Espanol2017,Yoshimoto2013,Li2014,Lavagnini2021,Trofimov2005,Groot1997} 
lies with the expression for the conservative force between 
particles, 
which includes attractive interactions and is derived from density dependent potentials. The most common form of the conservative force reads\cite{warren2003}
\begin{eqnarray}
	\bm{F}^C_{ij} =  A\omega^C(r_{ij})\bm{e}_{ij} + 
	B \left(\bar{\rho_i} + \bar{\rho_j} \right) \omega^d(r_{ij})\bm{e}_{ij},
	\label{eq2}
\end{eqnarray}
where $A<0$ and $B>0$ are the parameter strengths of the 
attractive and repulsive part of the potential, respectively.
Here, $r_{ij}$ indicates the distance between particles,
while $\bm{e}_{ij}$ is the unit vector from particle $i$ to $j$.
$\omega^C(r_{ij})$ and $\omega^d(r_{ij})$ are linear
weight functions, namely 
\begin{eqnarray}
	\omega^{C}(r_{ij}) = 
	\begin{cases}
		&1 - \frac{r_{ij}}{r_{c}}, \ \ r_{ij} \leq r_{c} \\
		& 0,  \  \ r_{ij} > r_{c},
	\end{cases} 
	\label{eq3}
\end{eqnarray}
where $r_c$ is a cutoff for the interactions, usually set to unity.
Also, $\omega^d(r_{ij})$ is defined by the same expressions as 
$\omega^{C}(r_{ij})$, but interactions have a smaller value than $r_c$,
namely $r_d=0.75$.

The repulsive term of the potential depends on the local 
neighborhood densities,
$\bar{\rho_i}$ and $\bar{\rho_j}$, and are given as follows:
\begin{eqnarray}
	\bar{\rho}_{i,j} = 
	\sum_{0<r_{ij}\le r_d}
	\frac{15}{2\pi r_d^3} \left( 1 - \frac{r_{ij}}{r_d}\right)^2,
	\label{eq4}
\end{eqnarray}
while, the random and dissipative forces are 
\begin{eqnarray}
	\bm{F}^D_{ij} = -\sigma \omega^D(r_{ij}) (\bm{e}_{ij} \cdot  \bm{v}_{ij})\bm{e}_{ij} ,
	\label{eq5}
\end{eqnarray}
\begin{eqnarray}
	\bm{F}^R_{ij} = \xi \omega^R(r_{ij}) \theta_{ij} \Delta t^{-1/2} \bm{e}_{ij},
	\label{eq26}
\end{eqnarray}
where $\sigma$ is the dissipative strength, $\xi$ the strength
of the random force, $\bm{v}_{ij}$ the relative velocity between particles, and $\theta_{ij}$ a random 
variable from a Gaussian distribution with unit variance.
The fluctuation--dissipation theorem dictates that 
$\sigma$ and $\xi$ be related by the following expression
\begin{eqnarray}
	\sigma = \frac{\xi ^2}{2 k_B T}.
	\label{eq7}
\end{eqnarray}
The weight functions for the random and dissipative forces are
\begin{eqnarray}
	\omega^D(r_{ij}) = \left[\omega^R(r_{ij})\right]^2 = 
	\begin{cases}
		\left( 1 - \frac{r_{ij}}{r_c}\right)^2,  & r_{ij} \leq r_{c} \\
		0,   & r_{ij} > r_{c}.
	\end{cases}
	\label{eq8}
\end{eqnarray}
Finally, system's temperature has been kept constant throughout the simulations
and set to unity (MDPD units). All simulations were done with a time-step $\Delta t = 0.01$ and dissipative strength $\sigma=4.5$.

The interaction levels and their respective $A_{ij}$
parameters for the non-bonded attractive 
interactions between particles and ions are presented
in \textbf{Table~\ref{tab:1}}
and have been obtained from our previous work with lipid membranes where parametrization was done based on water--octanol partitioning coefficients.\cite{Carnevale2024_MDPD_MARTINI, Kramarz2025} 
The bead labels have been renamed to follow the Martini convention for ease of comparison.
Here, these interactions are applied for the study of the 
aqueous systems laden with SDS surfactant and by setting $Q_d$ interactions equal to $Q_a$, while
$B=25$ remains constant due to the "no-go theorem" in MDPD.\cite{warren2013} The types
of beads and the groups they represent are illustrated
in \textbf{Figure~\ref{fig:model}} and are the same for both
the MD and MDPD models following the Martini mapping.

To build the SDS molecules harmonic bond and angle interactions are used, 
that is 
\begin{equation}
    E_{bond} = \frac{k}{2} \left( r_{ij} - r_0 \right)^2
\end{equation}
for the bonds with $k = 150$ and $r_0 = 0.5$, and 
\begin{equation}
    E_{angle} = \frac{k_A}{2} \left( \theta_{ijk} - \theta_0 \right)^2, 
\end{equation}
for each triad of consecutive particles with $k_A = 5$ and $\theta_0=180^{\circ}$. To compute long-range electrostatic interactions, 
the same pppm method was used. This method has been shown to be a suitable choice for MDPD simulations and does not present artificial ion pair formation.\cite{Hendrikse2024}

\begin{table}[t]
    \centering
    \begin{tabular}{c|cccccc}
    & P4  & Q0 & Qa   & Na    & C1 & C3 \\
    \hline
    P4   & I  & I & I & II  & VI & V \\ 
    Q0   & I  & IV & IV & II  & VI & V\\ 
    Qa   & I  & IV & IV & II  & VI & V\\ 
    Na   & II & II & II & III & VI & V \\ 
    C1 & VI  & VI & VI & VI & VI & VI \\ 
    C3 & V & V & V & V & VI & VI \\
    \end{tabular}
    \caption{MDPD-Martini-FF interaction matrix for parametrized MDPD beads 
    organized in six interaction levels (I-VI) with corresponding
    attractive parameters ($A_{ij}$):
$-50$ (I); $-43$ (II); $-34$ (III); $-30$ (IV); $-28$ (V); $-26$ (VI). These values were obtained by keeping the repulsive parameter constant for all interactions as $B=25$.}  
    \label{tab:1} 
\end{table}

\subsubsection{Mapping to real units}

To map MDPD reduced units to real units, we follow the original MDPD-Martini force field parametrization, which used the same scaling approach as in prior MDPD water parametrization studies at a coarse-graining level of three water molecules per bead.\cite{Ghoufi2011,Ghoufi2013}

All units conversions used are shown in Table \ref{tab:units}. In MDPD, the typical length scale is set by the attractive interaction cutoff $r_c$ and by the coarse-graining level. Its real value is obtained by comparing the density of beads in a simulation with the real molecular volume of water $V=30$~\AA.
In a similar manner, the density can be converted by using the molecular mass of water $M=18$~g/mol and Avogadro's number $N_A$. The energy scale in MDPD is set by the value of $k_BT$, with $k_B$ being the Boltzmann constant and $T$ the temperature.

Time scaling is obtained by conducting pure bulk water simulations and measuring the MDPD self-diffusivity, $D_{\rm MDPD}$, from the mean square displacement of the beads, which for the water model used is $D_{\rm MDPD}=0.049$.\cite{deng2025}
Using a time step of $\Delta t = 0.01$, the conversion shows this corresponds to 430 fs, which is 20 times longer than the 20 fs time step required in our MD simulations.
A comparison of the computational wall time between MD and MDPD can be found in the Supporting Information to this article.

\addtolength{\tabcolsep}{6pt} 
\begin{table}[b]
\centering
\caption{Conversion between MDPD and real units. The values presented are for the parametrization of water with coarse-graining level $N_m=3$. The MDPD values can be obtained from a bulk water simulation with $A=-50$, $B=25$, $r_c=1$, $r_d=0.75$. The length scale $r_c$ is converted by using the volume of a water molecule, that is, $V=30$ \AA. Time is scaled by the diffusion coefficient of water $D_{\rm water}$. The density conversion uses the molar mass of water $M$, and Avogadro's number $N_A$.} 
\begin{tabular}{llll}
  Parameter & MDPD & Conversion & MD  \\[1pt]
  \midrule
  $r_c$      & 1    & $(\rho N_m V)^{1/3} $  & 8.53 \AA     \\[5pt] 
  $\rho$     & 6.9  & $\rho N_m M/ N_Ar_c^3$ & 997 kg/m$^3$ \\[5pt] 
  $\gamma$   & 12.4 & $\gamma k_B T / r_c^2$ & 72 mN/m      \\[5pt]
  $\Delta t$ & 0.01 & $N_m D_{\rm MDPD} r_c^2 / D_{\rm water}$ & 0.43 ps   
\end{tabular}
\label{tab:units} 
\end{table}

\subsection{Systems}

We have carried out two different sets of simulations to compare different SDS properties. 
In the first case, aiming at determining interfacial properties, we have defined a simulation box with dimensions $L_x=L_y=17$~nm and $L_z=60$~nm. In the middle 
of the simulation box, we have placed a water slab with $17$~nm thickness with 
interface normals in the $z$-direction. Surfactant molecules were then placed 
on the interfaces, while an equal amount of counter-ions were placed in the water slab. A representation of such system is shown in \textbf{Figure~\ref{fig:surface-tension}a}.
Different initial surfactant surface coverage, $C$, were simulated with 
number of molecules ranging from $160$--$1440$ in the MDPD case, and from 
$160$--$960$ in the MD case. The surface coverage is defined as the total number of surfactants divided by the total interfacial area in the simulation, and 
for systems below the CAC, the surface coverage will be approximately equal to the surface excess concentration as most surfactants will stay at the interface.
This set of simulations was done on the 
canonical ensemble (NVT) and systems were allowed to equilibrate for $10^6$ time-steps while their energy was monitored to ensure they have reached an equilibrium (see Supporting Information), 
followed by $10^6$ time-steps production run, where surface tension was measured and averaged over time.
For MDPD simulations, the equations of motion in the NVT ensemble are integrated by a modified velocity-Verlet algorithm, which is the standard approach.\cite{Plimpton1995,Thompson2022,Pagonabarraga2001} 

In the second set of simulations, we have focused on SDS bulk properties.
All systems were composed of $1000$ surfactant molecules, $1000$ sodium 
ions and enough water beads to reproduce the desired concentration. Because the
water parametrization used in the development of the MDPD force-field follows a three-to-one mapping, the number of water beads in the MDPD simulations was higher than in the MD simulations, at a ratio of $4/3$. In a system with $45\%$ SDS, for example, we used 3889 and 5185 water beads in the MD and MDPD simulations, respectively. The number of water beads for different concentrations can be found in the Supporting Information. 
The molecules were placed in a random distribution inside a cubic 
simulation box and an initial energy minimization step was performed to avoid any numerical instabilities caused by possible overlap of beads in the initial configuration. The minimization was done by a conjugate gradient method with the tolerance set to $10^{-6}$. 
This set of simulations was done on the isothermal-isobaric (NPT) ensemble and ran for $3\times 10^6$ time-steps, which was enough to reach equilibrium based on measurements of the total energy of the system.
 To represent the NPT ensemble in MDPD, we chose to integrate the equations of motion using the isobaric-isoenthalpic (NPH) integrator in LAMMPS, which controls the pressure in the simulation, while temperature was still controlled by the MDPD thermostat, i.e. via the $\bm{F}^R$ and $\bm{F}^D$ forces.

\section{RESULTS AND DISCUSSION}

\subsection{Interfacial Properties}\label{subsec2}

\begin{figure}[t]
    \centering
    \includegraphics[width=0.5\linewidth]{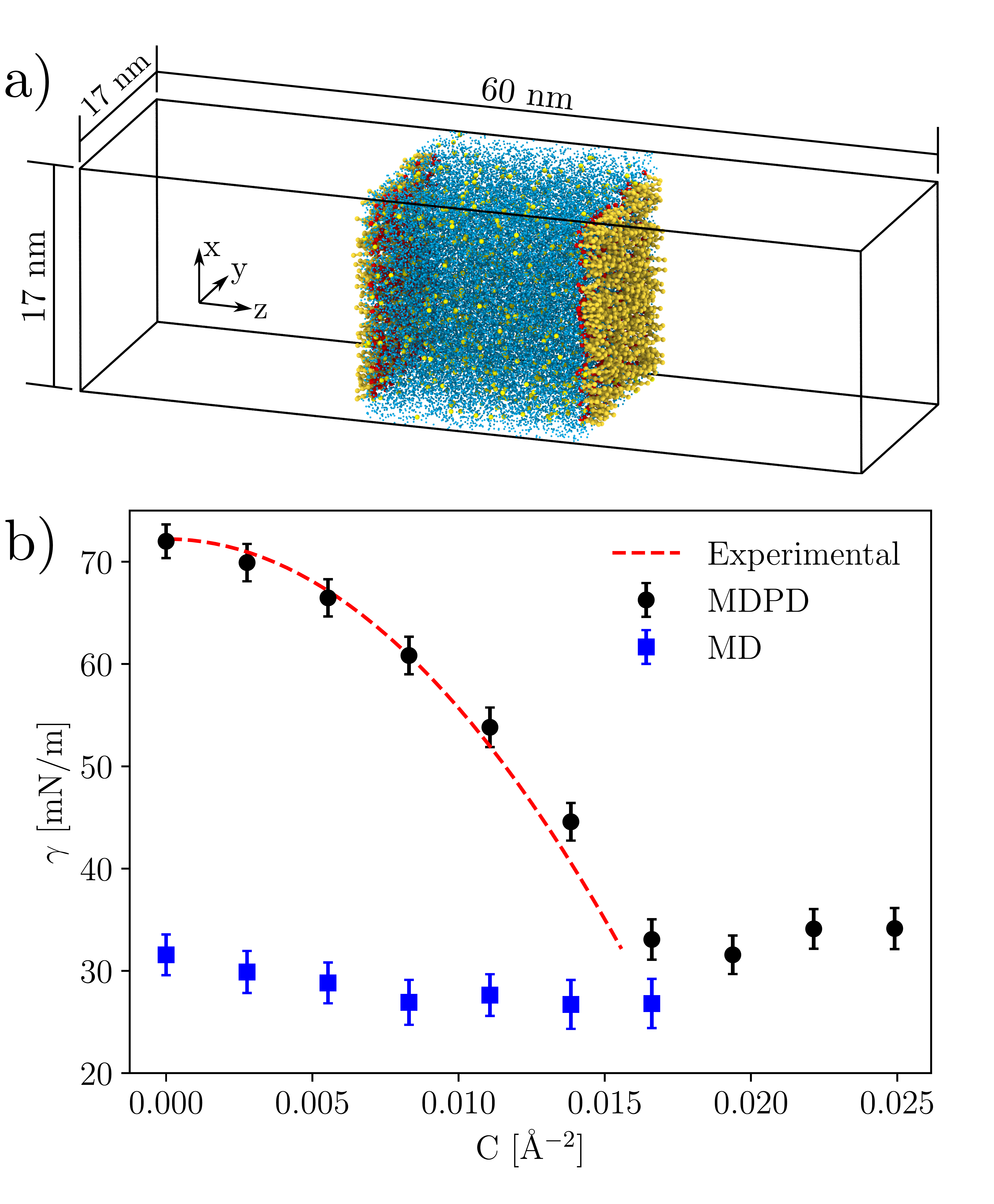}
    \caption{(a) Typical system setup for calculating
    surface tension in simulations. The same system parameters were used for
    both MD and MDPD simulations. Water beads are represented with a smaller radius for 
    visualization purposes. The color code is according to the model presented in \textbf{Figure~\ref{fig:model}}. 
    Snapshot has been obtained using Ovito software.\cite{Stukowski2009}
    (b) Surface-tension isotherms from MD and MDPD simulations, and
    experiment\cite{Fainerman2010} (see main text for details), as indicated. 
    The horizontal axis reflects the values of surface excess concentration, while
    $\gamma$ is surface tension.
    }
    \label{fig:surface-tension}
\end{figure}

The surface tension isotherm for the water--SDS--air interface can be calculated
in the simulations through the mechanical
route by using the setup of \textbf{Figure~\ref{fig:surface-tension}a.\cite{Kirkwood1949}} This is a relevant property as surfactants adsorb on the interface, reducing the interfacial tension and could change the dynamics of different phenomena, such as droplet coalescence and break-up.\cite{Arbabi_2023_soft_matter,Carnevale2024_phys_fluids}
Our simulations involve a slab of water laden with SDS molecules with periodic boundary conditions applied
in all directions of the simulation box. While the dimensions of the liquid slab coincide
with the size of the liquid phase in the directions parallel to the free surface ($x$ and $y$), 
the box size in the direction normal to the free surface
($z$, \textbf{Figure~\ref{fig:surface-tension}a}) is larger. 
The number of water beads added to the slab region was obtained from the average density of beads in bulk water simulations for both MD and MDPD.
Through this system setup, the pressure difference parallel and 
perpendicular to the flat, free surface (liquid--vapor surface)
would reflect the magnitude of the surface tension due to the presence 
of the liquid--vapor surface. 
This is expressed mathematically as follows:
\begin{equation} \label{eq:surface-tension}
    \gamma = \frac{L_z}{2}\left(P_{zz} - \frac{P_{xx} + P_{yy}}{2}\right),
\end{equation}
which in fact accounts for the presence of two liquid--vapor surfaces in the system and the
integration being throughout the simulation box ($L_z$) in the $z$ direction.
To compare these results with surface-tension isotherms obtained through
experiments,\cite{Fainerman2010} one needs to replace the bulk surfactant concentration
with the surface excess concentration, which is done by using the following 
relation\cite{Herdes2015}
\begin{equation} \label{eq:exp-surface-tension}
    C = -\frac{1}{RT}\frac{d\gamma}{d\ln c}.  
\end{equation}
Here, $R$ is the gas constant, and $c$ the bulk surfactant concentration. $\gamma$ is
surface tension, while $C$ the surface excess concentration. 
It should be noted that this relation should in principle hold for charge neutral
surfactant at a given constant temperature, $T$, and pressure, $P$, and
is valid in the pre-CAC region, for which the slopes $\frac{d\gamma}{d\ln c}$
can be calculated. Thus, one can produce experimental curves in the form obtained
by \textbf{Equation~\ref{eq:exp-surface-tension}} making a comparison between
simulation and experimental results possible.\cite{Herdes2015,Theodorakis2015Langmuir}
For a charged system, Equation 12 has a correction factor that depends on ionic activity coefficients. However because the bulk concentration of surfactants in the pre-CAC region is low and we have a 1:1 ratio of counter ions to surfactants, we can apply the usual Gibbs adsorption isotherm.\cite{martinez-balbuena2017} 
The isotherm obtained through the MDPD-Martini model
(\textbf{Figure~\ref{fig:surface-tension}}) shows a very good agreement with the 
experimental data, which are indicated by the red curve.\cite{Fainerman2010} 
In contrast, the MD-Martini model deviates considerably from both the
experimental and MDPD-Martini surface-tension isotherms. 
Moreover, the MD-Martini model exhibits significant deviations from 
the benchmark value of about 72~mN/m for water\cite{Fainerman2010,ianneti2024}. 
In addition, the surface tension at a concentration near the CAC is underestimated
in the MD-Martini model (\textbf{Figure~\ref{fig:surface-tension}}), 
in comparison with the value of about 35~mN/m found
in the case of the MDPD-Martini model and experiment.\cite{Fainerman2010}
Hence, the MDPD-Martini model seems to offer
advantages when investigating phenomena where surface tension is expected to
play an important role, such as droplet coalescence, \cite{Arbabi2023}
break-up,\cite{Carnevale2024} and oscillation phenomena.\cite{Ng2025}

\begin{figure}[t]
    \centering
    \includegraphics[width=1\linewidth]{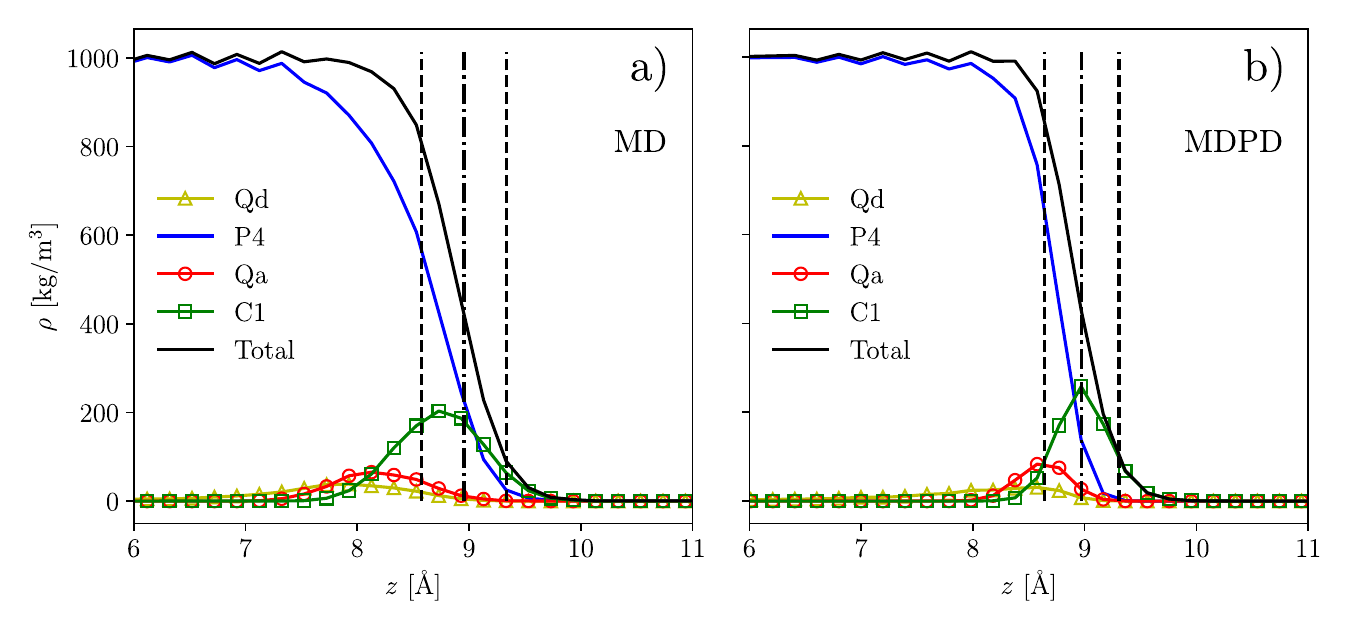}
    \caption{Density distribution of different beads across the liquid--vapor surface for
    (a) MD and (b) MDPD simulations. These distributions were obtained 
    for initial surface coverage of $C=0.0055$ \AA$^{-2}$,
    that is below the concentration at which micelles start to form for both methods.}
    \label{fig:densities}
\end{figure}

We have further explored the interfacial properties of the SDS--water systems
by calculating the density distribution of the various
chemical groups at the surface, juxtaposing the MD and MDPD results. 
In this case, computer simulation is particularly suitable for elucidating
the surface structure at the molecular resolution. 
While the MD-Martini and MDPD-Martini models are in good agreement,
slight differences can be observed. 
Considering a typical case for a specific surface coverage of  
$C=0.0055$~\AA$^{-2}$ (\textbf{Figure~\ref{fig:densities}}), we find
slightly broader distributions in the case of MD
than in the MDPD results, which may indicate that the
surfactant molecules in the MDPD model are more localized at the
liquid--vapor surface. This is also hinted by the fact that the 
water density-distribution shows slightly larger deviations from the
total density for all components in the case of MD simulations. 
These observations may indicate an overall more well-defined structure of
the interface in the case of MDPD-Martini models than in the MD-Martini model, 
despite the softest nature of interactions in the MDPD-Martini force-field.

\begin{figure}[t]
    \centering
    \includegraphics[width=0.5\linewidth]{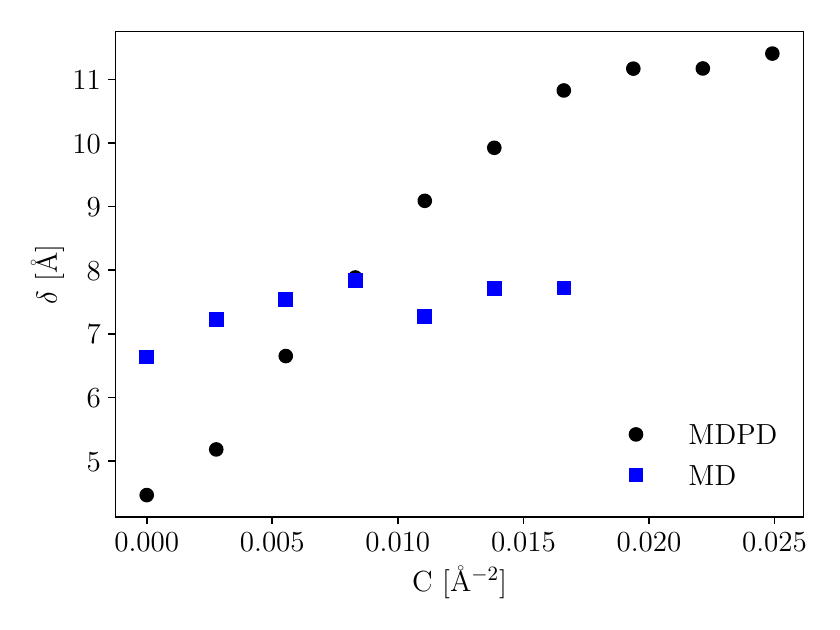}
    \caption{Thickness, $\delta$, of the liquid--vapor surface
    as a function of surface excess concentration for MD
    and MDPD methods, as indicated. $\delta$ is obtained
    through \textbf{Equation~\ref{eq:thickness}.} 
    }
    \label{fig:thickness}
\end{figure}

The sigmoidal shape of the total density as a function of the distance,
$z$ (\textbf{Figure~\ref{fig:densities}}), is often fitted to the following relation
\begin{equation}\label{eq:thickness}
    \rho(z) = \frac{\rho_l}{2} \left[ 1 - \tanh \left( \frac{2(z-z_0)}{\delta}\right) \right], 
\end{equation}
whence the thickness of the liquid--vapor surface, $\delta$, can
be obtained. 
Typically, it is expected that surfactant molecules will pack differently
as their concentration at the liquid--vapor surface increases, which
is manifested by a change in the slope when the thickness, $\delta$, is
plotted versus surface excess concentration, $C$.\cite{Carnevale2024,Hendrikse2023}
This effect clearly appears in the plot of \textbf{Figure~\ref{fig:thickness}}.
In the case of the MDPD-Martini model, initially, 
the rate of thickness increase is $405$~\AA$^{-1}$, 
which later decreases for higher values of the surface excess concentration
as the liquid--vapor surface saturates. The maximum thickness, $\delta$, recorded
is about 11~\text{\AA}. 
In contrast, the MD-Martini data indicate a smaller slope for
the thickness, which saturates to a maximum value, 
namely 7.5~\text{\AA}, at a lower surface
excess concentration in comparison with the MDPD data. 
Hence, the MD data obtained for the $\delta$ seem to be less sensitive to
the variation of surface excess concentration, in contrast to the data
based on the MDPD-Martini model.

% \begin{figure}
%     \centering
%     \includegraphics[width=0.7\linewidth]{figs/surfacetension.png}
%     \caption{Caption}
%     \label{fig:surfacetension}
% \end{figure}

% \begin{figure}
%     \centering
%     \includegraphics[width=0.7\linewidth]{figs/thickness.pdf}
%     \caption{Caption}
%     \label{fig:thickness}
% \end{figure}

\subsection{Bulk behavior}\label{subsec1}
We now turn our attention to various bulk properties of the system.
One property of interest here is the aggregation number,
$N_{\rm agg}$,\cite{Lavagnini2021} which can be calculated as follows
\begin{equation}
    N_{\rm agg} = \frac{\sum_{i} N_i^2P(N_i)}{\sum_{i} N_iP(N_i)}.
\end{equation}
$P(N)$ is the aggregation number distribution, while $N$ is the number of molecules
in a given micelle. This number is calculated when the distribution has
reached a constant value and the peak average value does not shift.
Moreover, the final value is calculated from three independent trajectories
following the procedure of previous MD-Martini simulations.\cite{Anogiannakis2020}
The monomers belonging to each micelle have been determined by applying 
a cluster algorithm with the distance chosen based on the 
Stillinger criterion.\cite{Stillinger1963} A histogram presenting the 
aggregation number distribution for both MD and MDPD simulations is available in the Supporting Information.
In the case of MD-Martini simulations, $N_{\rm agg} = 50 \pm 13$,
which is lower than expected for the given concentration  ($c=25\%$).\cite{Hammouda} 
The corresponding value in the case of the MDPD-Martini model is
$N_{\rm agg} = 55 \pm 21$. 
MDPD results showed a larger deviation from the average, but
the two values are in good agreement with each other
The underprediction of aggregation numbers from the standard
MD-Martini v2.2, which was used in this paper, is a known issue.\cite{wang2015,Peroukidis2021,peroukidis2022}
To properly reproduce the self-assembly process of SDS micelles,
modifications to the standard force-field or the use of 
slightly different force-field, such as having implicit solvent models, 
are usually necessary.\cite{Anogiannakis2020} 
It is worth noting that the more recent MD-Martini 
v3.0 is able to achieve results comparable to experimental values,\cite{Souza2021,coutinho2022} and we expect future versions of MDPD-Martini to be able to improve in the same direction as more interactions levels and beads are parametrized.

\begin{figure}[t]
    \centering
    \includegraphics[width=1\linewidth]{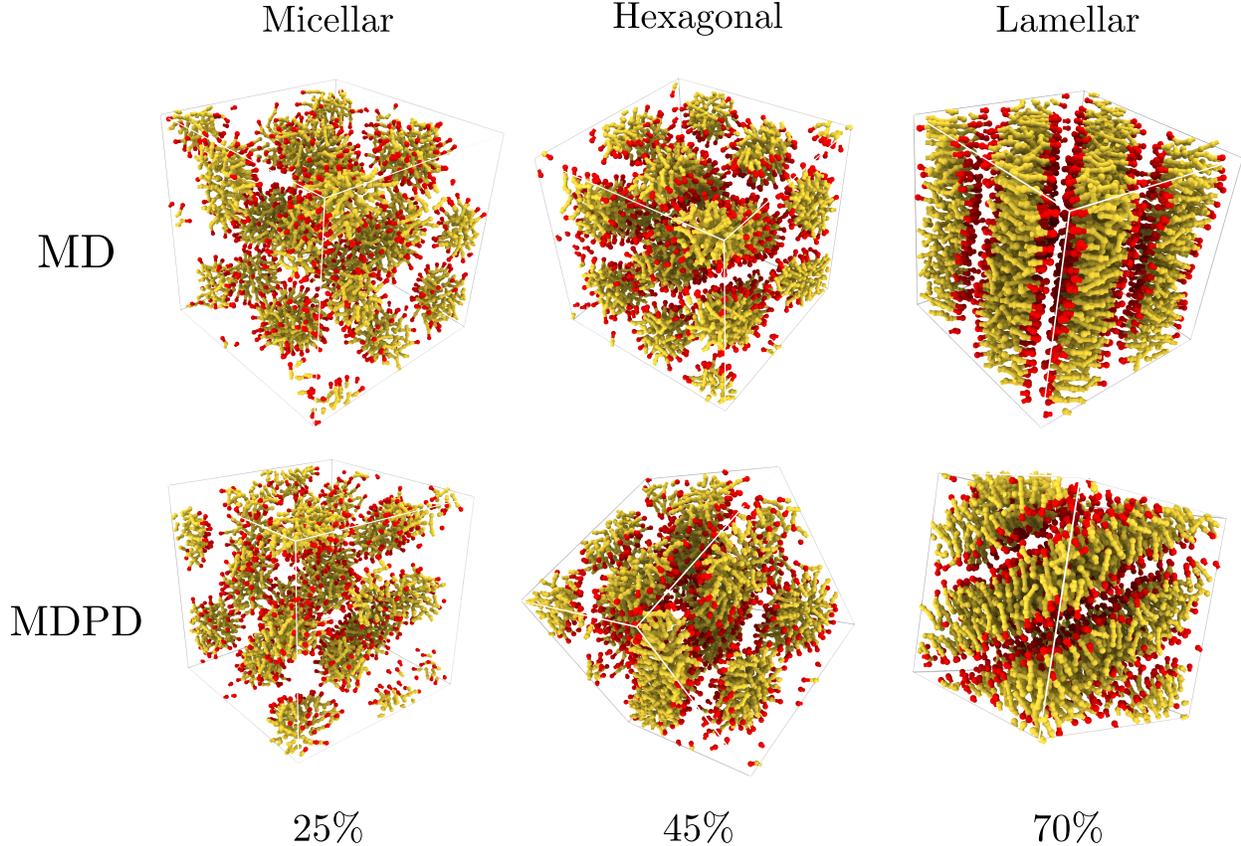}
    \caption{Different surfactant phases as a function of concentration. 
    Both MD and MDPD are capable of reproducing the same behavior, 
    in line with experiments.\cite{Fontell1981} 
    These simulations were done with 1000 SDS molecules, 1000 counter ions, 
    while the number water beads was adjusted for the desired concentration. 
    Only surfactants are shown for better visualization of the morphologies.
    Snapshots have been obtained by using OVITO.\cite{Stukowski2009}
    }
    \label{fig:phases}
\end{figure}

The next property that we have examined is
the morphology of various SDS--water systems
for different surfactant concentration above the CAC.
The expected morphologies for the SDS--water system
are micelles, hexagonal and lamellar.\cite{Fontell1981}
These have been confirmed by MD-Martini simulations\cite{Anogiannakis2020} 
and we have also reproduced here the MD-Martini results
for the same concentrations, which are presented
in \textbf{Figure~\ref{fig:phases}}. 
The MDPD-Martini model is able to reproduce these morphologies
in line with the MD-Martini\cite{Anogiannakis2020} results and 
experiments.\cite{Fontell1981} 
In \textbf{Figure~\ref{fig:phases}}, the number
of micelles are 18 for MDPD versus 20 in MD. 
The average distance between the cylinders in the hexagonal phase is about
43.4~\AA ~in MDPD versus 39.8~\AA ~in MD, while
the lamellar width is 31.8~\AA ~in MDPD versus
28.7~\AA ~in MD.

\subsection{Coherent Scattered Intensity}

\begin{figure}[t]
    \centering
    \includegraphics[width=.6\linewidth]{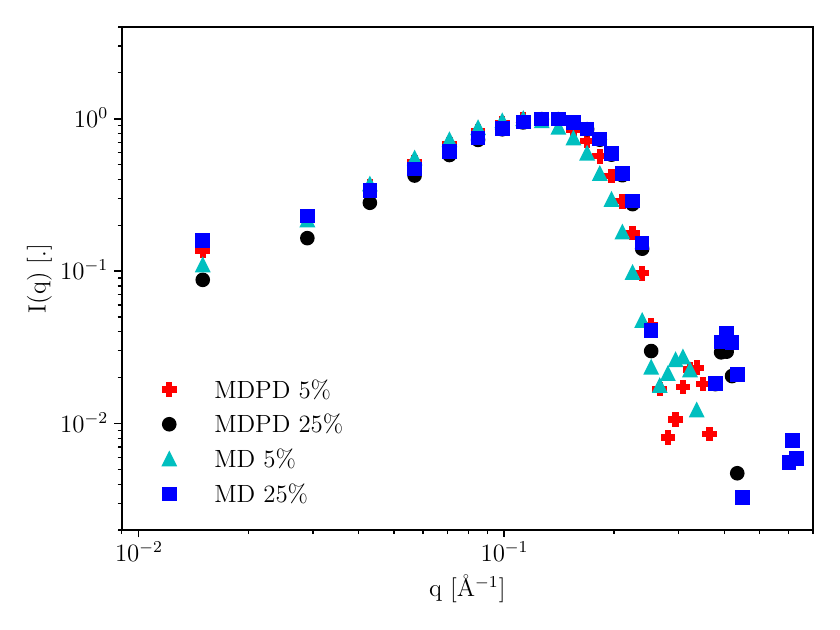}
    \caption{Coherent scattered intensity versus wavenumber, $q$.
    The wavelength can be obtained through the expression $\lambda = 2\pi/q$.
    Both MD and MDPD present the same peaks for their respective concentrations. 
    For $5\%$ $\lambda\approx5.7$~nm, while for $25\%$, $\lambda\approx4.5$~nm}
    \label{fig:scatter}
\end{figure}

We have examined the scattering pattern of an
SDS solution at 5\% and 25\%, which can be compared with
SANS experimental results.\cite{Hammouda} 
In the simulations, this can be obtained through the
coherent scattered intensity from the Fourier Transform of
the radial distribution functions $g_{ij}(r)$
expressed as follows
\begin{equation} \label{eq:scattering}
    I(q) = \sum_{i=1}^n\sum_{j=1}^n x_ix_jf_if_j\left(1 + \rho\int_0^{\infty}
    4\pi r^2 (g_{ij}(r)-1)\frac{\sin(qr)}{qr}dr \right),
\end{equation}
where $n$ is the number of different bead types, 
$x_i$ is the concentration of 
beads of type \textit{i}, 
$f_{i,j}$ are the atomic form factors and $g_{ij}(r)$
is the radial distribution function between beads 
of type \textit{i} and \textit{j}.

To allow for a direct comparison with previous data,\cite{Anogiannakis2020}
which are also here reproduced for different surfactant concentration in 
the micelle regime, and take into account the coarse-grained nature of
the MD and MDPD models, we set $f_i=f_j=1$. The coherent scattered 
intensity for solutions with different surfactant concentration as
indicated is presented in \textbf{Figure~\ref{fig:scatter}}.
The MD and MDPD data are in very good agreement for both
concentrations shown here. 
The position of the peak approximately reflects the mean value
of the distance between the micelles, which decreases as the
surfactant concentration increases.\cite{Hendrikse2024}
These distances correspond to
$\lambda \approx 5.7~\rm nm$ for 5\% surfactant concentration,
while $\lambda \approx 4.5~\rm nm$ for 25\% concentration, 
where $\lambda=2 \pi / q$.
The experimental values are 6.9~nm for 20\% concentration
and 8.6~nm for 5\% at temperature of 293~K. 
Hence, both MD and MDPD data generally underpredict the 
wavelength of the the peak intensity. 
This difference is due to the lower 
aggregation numbers obtained in our simulations. 
A relation between the aggregation number and the mean micelle distance
can be obtained from the intermicellar coulombic repulsion model (ICRM)\cite{Patist2001} and is given by $N_{agg}=cN_A\lambda^3$, where
$c$ is the molar concentration and $N_A$ is Avogadro's number. Our results
match with the estimates from this equation for both concentrations.
In addition, the shapes of the curves for both MD and MDPD simulations is the same.
In comparison with experimental data,\cite{Hammouda}
certain differences are expected
at high wavelengths, due to the coarse-grained character of the 
simulation models, as has been discussed before.\cite{Anogiannakis2020}

\section{CONCLUSIONS}
In this study, we have presented the MDPD-Martini model for SDS
surfactant, which is capable of describing several properties of
water--SDS systems. Moreover, a direct comparison between the MDPD-Martini
and MD-Martini models has been conducted with results for 
MD simulations from the literature\cite{Anogiannakis2020}
reproduced in our study. 
Our analysis has indicated that the MD-Martini and MDPD-Martini models
are in very good agreement with each other and with experimental results.
Moreover, the MDPD-Martini model is better able to reproduce the surface
tension isotherm, in line with the experimental data.\cite{Fainerman2010}
Results are also in line with a previous 
MDPD model for SDS surfactant.\cite{Hendrikse2024}

The MDPD-Martini model presented here relies on transferable 
interactions\cite{Carnevale2024_MDPD_MARTINI,Kramarz2025}
and explicitly modeling the cations as point charges as separate
beads, an approach which has been only recently employed
in MDPD models.\cite{Hendrikse2024} A current limitation of our force-field is the small number of parametrized beads and interaction levels, which reduce the scope of chemical specificity and the number of possible molecule representations. However,
further steps in the development of the general-purpose MDPD-Martini will 
include additions to the current interaction matrix based on the
Martini mapping,\cite{Souza2021} as well as
the integration of methodologies, such as those 
related to reactivity\cite{Sami2023} and 
the G\={o}Martini approach.\cite{Poma2017}
In view of the computational speed-up provided by the 
MDPD-Martini models in comparison with MD,\cite{Carnevale2024_MDPD_MARTINI}
we anticipate that the current study opens opportunities for
further applications of the MDPD-Martini models to a greater range
of soft matter systems, including systems that require the 
explicit modeling of charges. Finally, our study can provide 
further inspiration for the development of general-purpose
force-field based on non-LJ interactions.

\begin{acknowledgement}
This research has been supported by the National 
Science Centre, Poland, under
Grant No.\ 2019/34/E/ST3/00232. 
We gratefully acknowledge Polish high-performance computing infrastructure 
PLGrid (HPC Centers: ACK Cyfronet AGH) for providing computer facilities and 
support within computational grant no. PLG/2025/018743.
\end{acknowledgement}

\begin{suppinfo}
Additional figures on energy equilibration, micelle aggregation, surface tension measurements, and tables on systems description and computational time comparisons.
\end{suppinfo}

\bibliography{bib}

\end{document}